\documentclass[twocolumn]{aastex63} %trackchanges
\usepackage[capbesideposition=right]{floatrow}
\usepackage{wrapfig}
\usepackage[right = 0.5in, bottom = 1in]{geometry}
\usepackage{hyperref}

\newcommand{\Rs}{$ R_{\odot} $}
\def\ion[#1 #2]{#1\,{\sc #2}}

\submitjournal{ApJ}
\shortauthors{Boe et al.}

\begin{document}

\title{The Solar Minimum Eclipse of 2019 July 2. III. Inferring the Coronal $T_e$ with a Radiative Differential Emission Measure Inversion}

\author{Benjamin Boe}
\affil{Institute for Astronomy, University of Hawaii, Honolulu, HI 96822, USA}

\author{Cooper Downs}
\affil{Predictive Science Inc., San Diego, CA, 92121, USA}

\author{Shadia Habbal}
\affil{Institute for Astronomy, University of Hawaii, Honolulu, HI 96822, USA}

\correspondingauthor{Benjamin Boe}
\email{bboe@hawaii.edu}
\begin{abstract}
Differential Emission Measure (DEM) inversion methods use the brightness of a set of emission lines to infer the line-of-sight (LOS) distribution of the electron temperature ($T_e$) in the corona. DEM inversions have been traditionally performed with collisionally excited lines at wavelengths in the Extreme Ultraviolet (EUV) and X-ray. However, such emission is difficult to observe beyond the inner corona (1.5 \Rs), particularly in coronal holes. Given the importance of the $T_e$ distribution in the corona for exploring the viability of different heating processes, we introduce an analog of the DEM specifically for radiatively excited coronal emission lines, such as those observed during total solar eclipses (TSEs) and with coronagraphs. This Radiative DEM (R-DEM) inversion utilizes visible and infrared emission lines which are excited by photospheric radiation out to at least 3 \Rs. Specifically, we use the \ion[Fe x] (637 nm), \ion[Fe xi] (789 nm), and \ion[Fe xiv] (530 nm) coronal emission lines observed during the 2019 July 2 TSE near solar minimum. We find that despite a large $T_e$ spread in the inner corona, the distribution converges to an almost isothermal yet bimodal distribution beyond 1.4 \Rs, with $T_e$ ranging from 1.1 to 1.4 in coronal holes, and from 1.4 to 1.65 MK in quiescent streamers. Application of the R-DEM inversion to the Predictive Science Inc. magnetohydrodynamic (MHD) simulation for the 2019 eclipse validates the R-DEM method and yields a similar LOS Te distribution to the eclipse data. 
\end{abstract} 

\section{Introduction} 
\label{intro}
Differential Emission Measure (DEM) inversion techniques combine observations of multiple emission lines to infer the line-of-sight (LOS) distribution of the electron temperature ($T_e$) in an optically thin plasma, such as the solar corona. DEM inversions are commonly used to explore coronal thermodynamics with emission lines at X-Ray (e.g. \citealt{Pottasch1964, Kepa2006, Kepa2022}) and Extreme Ultra Violet (EUV; e.g. \citealt{Withbroe1978, Hannah2012, Cheng2012, Cheung2015}) wavelengths. EUV DEMs are a useful tool for {inferring $T_e$ in} active regions \citep{Aschwanden2015}, and the solar cycle dependence of $T_e$ in the corona \citep{Morgan2017}. X-ray DEMs have even been used to study the temperature of stellar coronae \citep{Gudel2001}.
\par

A significant limitation of DEM inversion analysis based on X-Ray and EUV lines is that the lines are predominantly collisionally excited, which means the emission is proportional to the density squared, and so they decline in brightness rapidly at larger helioprojective distances. There can be radiative excitation of EUV lines but that requires flux from the same collisionally excited line lower down in the corona (the solar disk does not emit any meaningful amount of EUV photons). Recent developments have proven that EUV lines can be detected out to 2 \Rs \ with \textit{Proba-2}/SWAP \citep{Goryaev2014} and 2.5 \Rs \ with \textit {GOES}/SUVI \citep{Seaton2021}, but the EUV lines are difficult to observe beyond 1.5 \Rs, and the relative contribution of resonant scattering is not as well characterized since its source brightness {(i.e., emission from the low corona)} has spatial and temporal variation. The visible and infrared coronal lines, on the other hand, can be radiatively excited by photospheric emission and in turn, can often be observed at helioprojective distances of up to 3.4 \Rs \ or more (see \citealt{Boe2022}).

\par
Another drawback of DEM inversion performed specifically with bandpasses in the EUV is that multiple lines are present in each bandpass and so the observed brightness leads to complex temperature response functions (e.g., \citealt{Odwyer2010, Boerner2012}). Solving a DEM with EUV observations is therefore rather challenging; it requires non-trivial inversion methods (e.g. \citealt{Hannah2012, Guennou2012a, Guennou2012b, Aschwanden2015, Morgan2019}) and clever techniques to reduce the computationally expensive operation (e.g., \citealt{Kashyap1998, Plowman2013, Cheung2015, Pickering2019}). (See \cite{DelZanna2018b} for a detailed overview of the history of DEM inversion methods and applications) 
 
\par
{
Outside of EUV and X-ray DEM inversions, the coronal electron temperature has been inferred with emission line ratios at wavelengths in the ultraviolet from space (e.g., \citealt{Raymond1997,Ko2002}) and in the visible and infrared during total solar eclipses (e.g., \citealt{Boe2020a, Boe2022, DelZanna2023}). Without emission lines, the effective temperature has been inferred using the density profile inferred from polarized brightness measurements (e.g., \citealt{Munro1977}). However, the effective temperature is a combination of the electron and proton temperatures, which will follow significantly different profiles (see \citealt{Esser1997}). Isolating the electron and proton temperatures using continuum-based analysis is thus not possible without other data or assumptions (e.g., \citealt{Doyle1999}). The electron temperature of the corona can also be deduced from in situ measurements of ionic composition (e.g., \citealt{Ko1996, Ko1997, Smith2003, Habbal2010a, Landi2014}) since the ions will cease charge exchange processes at some distance from the Sun (i.e., they freeze-in their ionic state; see \citealt{Boe2018, Gilly2020}). Unfortunately, any in situ driven analysis lacks precise information about the coronal structure (and freeze-in distance) that each solar wind parcel originated from. None of these methods have provided a complete survey of the distribution of the electron temperature throughout the corona, and in all but a few cases are only able to probe very small portions of the corona. Only visible and infrared emission line observations have shown the potential to quantify the LOS $T_e$ distribution throughout the entire middle corona from below 1.5 \Rs to perhaps as much as 6 \Rs.
}
\par

In this paper, we implement a new Radiative-DEM method, which is introduced for lines that are radiatively excited (see Section \ref{RDEM}). The R-DEM is applied to observations of the \ion[Fe x] (637.5 nm), \ion[Fe xi] (789.2 nm), and \ion[Fe xiv] (530.3 nm) emission lines from the 2019 total solar eclipse (TSE; see Section \ref{Eclipse}). We also implement the R-DEM method on forward-modeled line emission from the Predictive Science Inc. (PSI) Magnetohydrodynamic (MHD) simulation (see Section \ref{PSI}) and compare the inversion results to the actual $T_e$ distribution along the LOS in the MHD model. We discuss the details of the findings, including comparisons between the eclipse and model inferences and actual LOS $T_e$ distribution in Section \ref{disc}. Concluding remarks are given in Section \ref{conc}.

% \newpage
\section{2019 July 2 Total Solar Eclipse}
\label{Eclipse}
\begin{figure*}[t!]
\centering
\includegraphics[width = \textwidth]{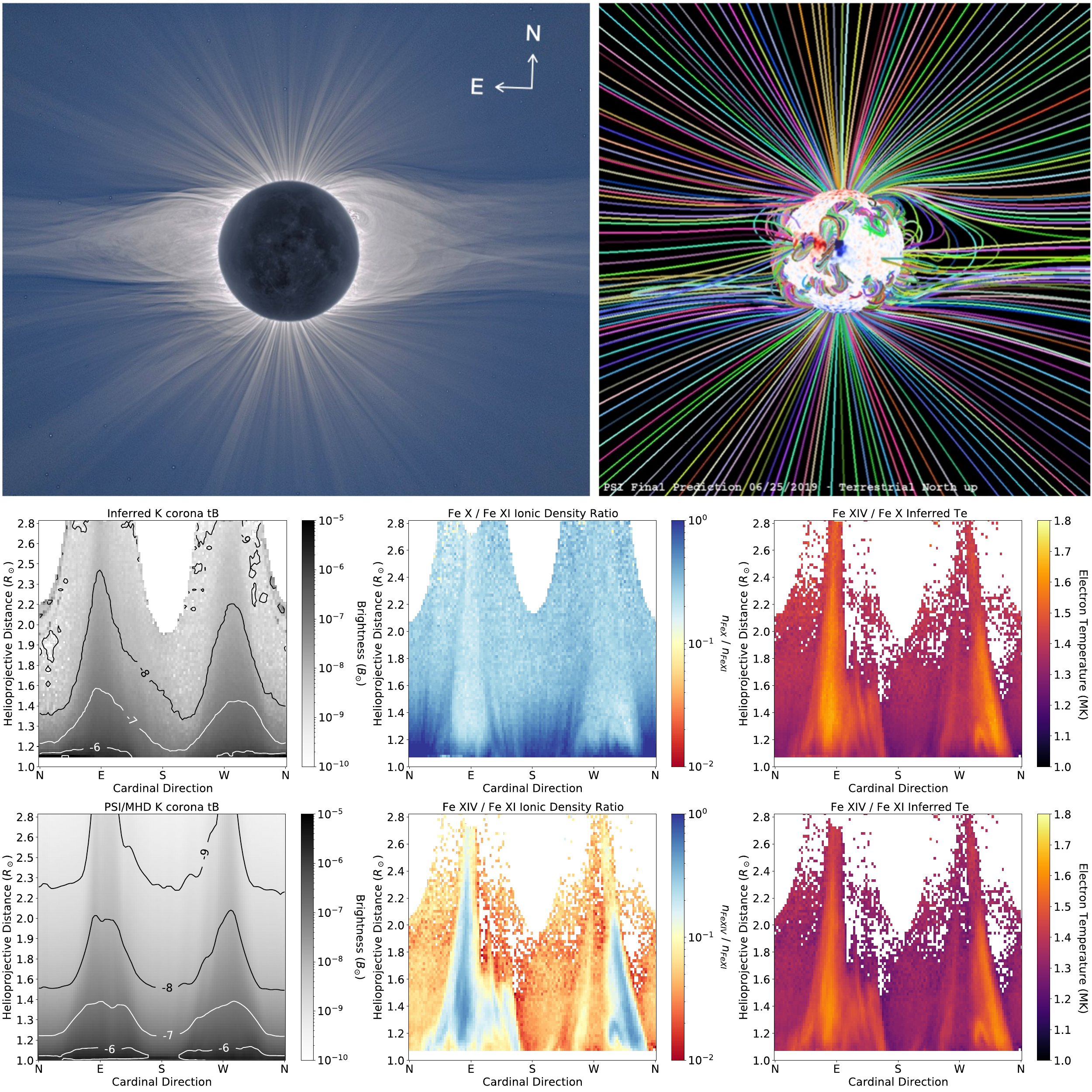}
\caption{Top left: White light image of the 2019 TSE (first presented in \citealt{Boe2020b}). Top right: PSI MHD model field line prediction of the eclipse (see Section \ref{PSI}). The bottom two rows contain a collection of inferred (and modeled) coronal properties in a cartesian representation of polar coordinates from \cite{Boe2021a, Boe2022}. Bottom left: K corona brightness from the eclipse (middle) and PSI MHD model (bottom). Bottom middle: Inferred ionic density ratio of \ion[Fe x]/\ion[Fe xi] (middle) and \ion[Fe xiv]/\ion[Fe xi] (bottom). Bottom right: Inferred $T_e$ from the \ion[Fe xiv]/\ion[Fe x] (middle) and \ion[Fe xiv]/\ion[Fe xi] (bottom) ionic density ratios. \vspace{2mm}}
\label{Fig1}
\end{figure*}

To explore the $T_e$ distribution in the corona with a new Radiative-DEM inversion procedure (see Section \ref{RDEM}), we use visible observations of the brightness of \ion[Fe x] (637.5 nm), \ion[Fe xi] (789.2 nm) and \ion[Fe xiv] (530.3 nm), which were acquired during the 2019 July 2 TSE in Rodeo, Argentina. This eclipse occurred very close to solar minimum when the corona was dominated by a dipolar field component, as showcased by the white light eclipse image and the magnetic field line extrapolation of the PSI MHD model (see Section \ref{PSI}) in the top panels of Figure \ref{Fig1}. The brightness of the emission lines was acquired with narrowband telescope systems that have been deployed at a number of recent TSEs (see \citealt{Habbal2021, Boe2020a, Boe2018}).

\par
\begin{figure}%[t!]
\centering
\includegraphics[width = 3 in]{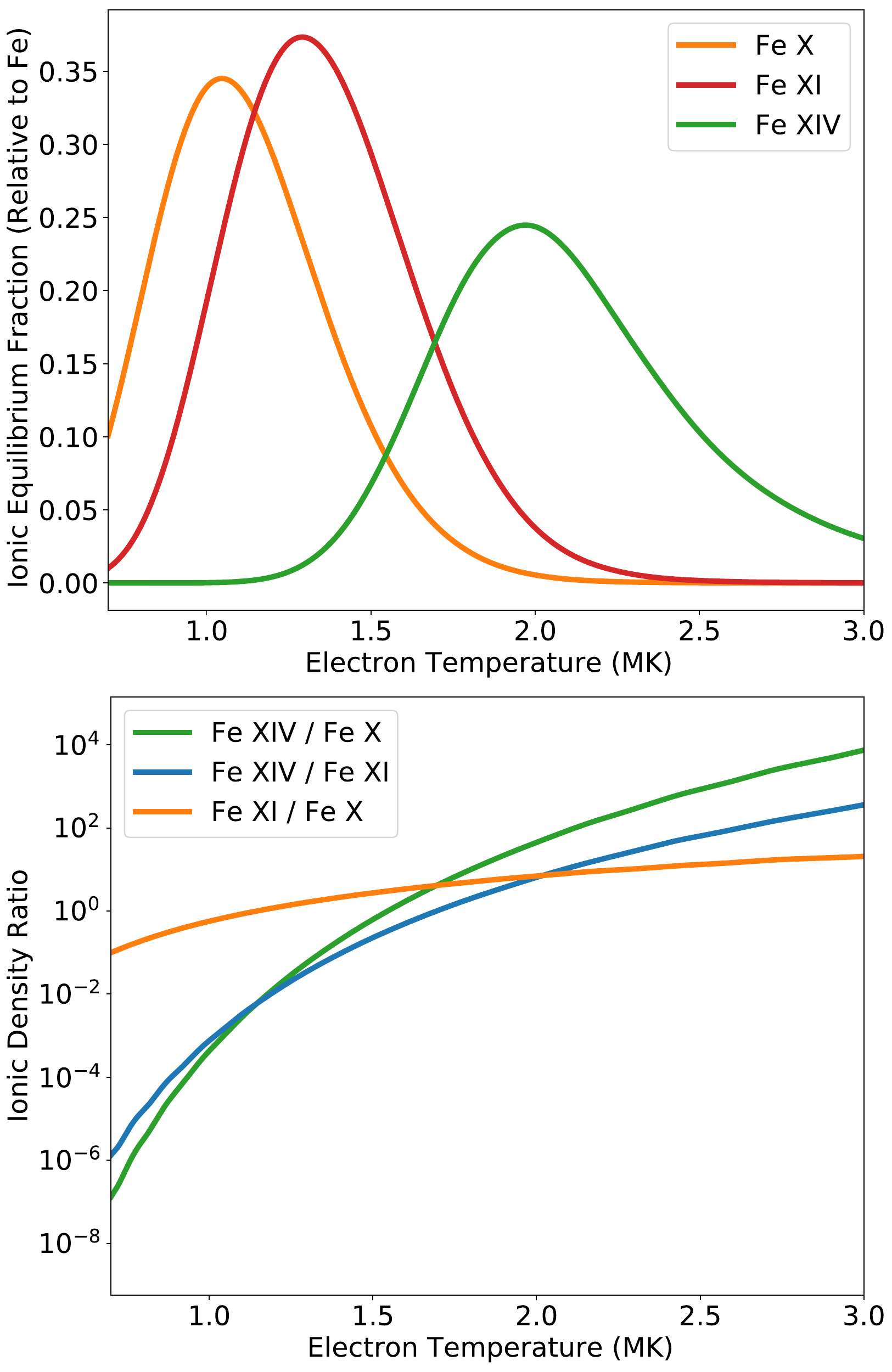} %\textwidth for two column
\caption{Top: Ionic equilibrium curves of \ion[Fe x], \ion[Fe xi], and \ion[Fe xiv] versus $T_e$ with values interpolated from CHIANTI. Bottom: Ionic equilibrium density ratios vs $T_e$ from the curve above. \vspace{2mm}}
\label{Fig2}
\end{figure}

The temperature responses of the \ion[Fe x], \ion[Fe xi], and \ion[Fe xiv] lines span coronal electron temperatures ($T_e$) from 0.8 to 2.5 MK. The ionic equilibrium curves that we use in this work are shown in the top panel of Figure \ref{Fig2}. The curves are generated from version 10 of the CHIANTI database \citep{Dere1997, DelZanna2021}, which have been interpolated from their recorded spacing of $\Delta$ log(K) = 0.05. Higher $T_e$ plasma can exist in active regions below 1.2 \Rs, but the vast majority of coronal plasmas that escape into the solar wind are found to have a coronal temperature between 1 and 2 MK \citep{Habbal2021}. Further, this eclipse occurred near solar minimum when there were no active regions on the Sun. 

\par
To isolate the line emission, we use one telescope with a bandpass centered on the wavelength of the emission line, and another {about 1 to 3 nm away from} the line to observe the background continuum brightness. The set of continuum observations from this eclipse were used by \cite{Boe2021a} (Paper I) to study the K (electron) and F corona (dust) brightness using a novel color-based inference method. The calibration for the continuum data was leveraged off the Mauna Loa Solar Observatory's (MLSO) K-coronagraph (K-Cor)\footnote{K-Cor DOI: 10.5065/D69G5JV8; \url{https://mlso.hao.ucar.edu/mlso_data_calendar.php}} polarized brightness data from the same day as the eclipse. The inferred K corona brightness from Paper I can be found in the middle left panel of Figure \ref{Fig1}. \cite{Boe2022} (Paper II) then extended the calibration to the simultaneous line emission observations to isolate the brightness of \ion[Fe x], \ion[Fe xi], and \ion[Fe xiv].

\par
In Paper II, we inferred the LOS integrated ionic column density ratio which we will use in this work to perform the R-DEM inversion. Specifically, we isolated the radiative component of the line emission observations by removing the fraction of brightness that originated from collisional excitation based on the PSI MHD model (see Section \ref{PSI}). The radiative line ratios were then used to infer the integrated column density ratios of \ion[Fe x]/\ion[Fe xi] and \ion[Fe xiv]/\ion[Fe xi] after accounting for the atomic radiation processes of each ion as well as the effect of variable line-widths integrated over each bandpass at different helioprojective distances (based on previous observations). Note that since these ratios are integrated along the same LOS, they are equivalent both to the column density ratio as well as the average volumetric density ratio along each LOS. 
\par
The final inferred ionic density ratios are shown in the bottom middle panels of Figure \ref{Fig1}. These ionic density ratio maps clearly show that \ion[Fe xi] is the most abundant ion of the three. \ion[Fe x] has a similar density in the low corona (below 1.3 \Rs) but declines relative to \ion[Fe xi] at higher helioprojective distances. As for the \ion[Fe xiv] ion, it has a similar density to \ion[Fe xi] in the cores of the coronal streamers but is otherwise significantly less abundant than \ion[Fe xi]. Paper II then took these ionic equilibrium density ratios and inferred $T_e$ in the corona, using the correspondence between the ionic equilibrium and $T_e$ (i.e., the curves shown in the bottom panel of Figure \ref{Fig2}). The \ion[Fe xi]/\ion[Fe x] ratio in particular did not return a realistic average $T_e$, likely due to the very shallow slope of the density ratio vs $T_e$ curve. The other two $T_e$ inferences, from the \ion[Fe xiv]/\ion[Fe x] and \ion[Fe xiv]/\ion[Fe xi] ionic density ratios, are shown in the bottom right panels of Figure \ref{Fig2}. In this work, we implement a new Radiative-DEM procedure that uses the corrected \ion[Fe x]/\ion[Fe xi] and \ion[Fe xiv]/\ion[Fe xi] inferred density ratios to infer the LOS distribution of $T_e$ in the corona.

\section{Radiative Differential Emission Measure}
\label{RDEM}
\subsection{R-DEM Formalism}
\label{formalism}
The traditional DEM is defined as (see \citealt{Withbroe1978, DelZanna2018b}):
\begin{eqnarray}
\label{DEM}
DEM(T_e) = N_e^2 \frac{dh}{dT_e},
\end{eqnarray}
where $N_e$ is the electron density (squared since it is causing collisional excitation of the ions). The DEM represents the density squared weighted $T_e$ distribution along a given LOS, in units of cm$^{-5}$ K$^{-1}$. 
\par

The brightness, $\beta_i$, of an emission line $i$ from an ion of element $Z$ with elemental abundance $Ab(Z)$, can be related to the integral of the temperature over the contribution function, $C_i(T_e, N_e, \lambda_i)$, multiplied with the DEM function. The brightness is then:
\begin{eqnarray}
\label{DEMline}
\beta_{i} = Ab(Z) \int_{T_e}  C_i(T_e, N_e, \lambda_i) \ DEM(T_e) \ dT_e.
\end{eqnarray}
\par
The traditional DEM formulation only works for collisionally excited lines, which are the main type of line emission one can usually expect at X-ray and EUV wavelengths (see Section \ref{intro}). The contribution function is thus a complicated interdependence of the nature of the given ionic line transition, the electron density (since the lines are collisionally exited), and the temperature dependence of the ionic species. 
\par
For the visible lines observed during the eclipse, on the other hand, the lines are almost entirely radiatively excited and so a different formalism is required. We {thus} introduce a slightly modified Radiative-DEM (R-DEM), which will replace the {$DEM(T_e)$} term in Equation \ref{DEMline}, where
\begin{eqnarray}
\label{RDEM}
RDEM(T) = N_e \frac{dh}{dT}, %\mbox{--}
\end{eqnarray}
which will have units of cm$^{-2}$ K$^{-1}$. Unlike the collisional DEM inversion, the R-DEM inversion has a significantly simpler contribution function {as it does not depend on the electron density (i.e., $C_i(T_e, \lambda_i$)). Rather, it is} only a combination of the straightforward radiative ionic transition parameters based on radiative excitation from the photospheric emission combined with the temperature response of the ion (i.e. the curves in the top panel of Fig. \ref{Fig2}). The brightness of a radiatively excited line is then:

\begin{eqnarray}
\label{RDEMline}
\beta_{i} = Ab(Z) \int_{T_e}  C_i(T_e, \lambda_i) \ RDEM(T_e) \ dT_e.
\end{eqnarray}
One can also calculate the Radiative Emission Measure, or R-EM, as the integral of the R-DEM over a defined temperature range of $\Delta T$ {centered on temperature $T_0$},
\begin{eqnarray}
\label{REM}
REM(T) = \int_{T_0 - \frac{\Delta T}{2}}^{T_0 + \frac{\Delta T}{2}} RDEM(T) \ dT,
\end{eqnarray}
which represents the amount of emitting plasma at the specified temperature range in units of cm$^{-2}$.
\par
Next, we simplify the inversion procedure given the particular properties of radiative excitation by using calibrated ionic column density ratios, rather than the absolute emissivity. In Paper II, we used the ratio of the brightness between two radiative lines to infer the relative ionic density with the following equation, 
\begin{eqnarray}
\label{eqn2Line}
\frac{n_{j}}{n_{k}} = \frac{\beta_j \ A_{k}  \ g_{l,j} \ g_{u,k} \ {\nu_{k}}^3}{\beta_k \ A_{j} \ g_{u,j} \ g_{l,k} \ {\nu_{j}}^3},
\end{eqnarray}
where, for ions $j$ and $k$, $n$ is the number density, $\nu$ is the frequency of the line emission, $A$ is the Einstein coefficient for spontaneous emission, and $g$ is the statistical weight for the given energy level. {Equation \ref{eqn2Line}} assumes that the observed brightness $\beta$ has been calibrated in solar disk units (i.e., B$_\odot$), thus removing the spectral dependency on the photospheric spectrum (see \citealt{Boe2020a, Boe2022}). It also assumes the brightness to be caused entirely by radiative excitation. The ionic density ratios as computed in Paper II already removed the collisional excitation component from the brightness using the ratio of the collisional and radiative excitation predicted by the PSI MHD simulation (which is only important in the streamers below $\approx 1.5$ \Rs). The calibrated ionic density ratios inferred in Paper II using Equation \ref{eqn2Line} for \ion[Fe x]/\ion[Fe xi] and \ion[Fe xiv]/\ion[Fe xi] are shown in the bottom middle panels of Figure \ref{Fig1}.

\par
To express the R-DEM inversion in terms of the ionic density ratio, we take the ratio of the R-DEM brightness definition from Equation \ref{RDEMline} and combine it with Equation \ref{eqn2Line}, giving, 
\begin{eqnarray*}
\frac{\beta_j}{\beta_k} &&= \frac{n_{j}\ A_{j} \ g_{u,j} \ g_{l,k} \ {\nu_{j}}^3}{n_{k} \ A_{k}  \ g_{l,j} \ g_{u,k} \ {\nu_{k}}^3} \\
&&= \frac{  Ab(Z_j) \int C_j(T_e, \lambda_j) \ RDEM(T_e) \ dT_e} {Ab(Z_i) \int  C_kT_e, \lambda_k) \ RDEM(T_e) \ dT_e},
\end{eqnarray*}
which reduces to,
\begin{eqnarray}
\label{eqnFinalRDEM}
 \frac{n_{j}} {n_{k}} =  \frac{ Ab(Z_j) \int C'_j(T_e) \ RDEM(T_e) \ dT_e} {Ab(Z_k) \int  C'_k(T_e) \ RDEM(T_e) \ dT_e}.
\end{eqnarray}
\par
The new contribution function, $C'(T_e)$, has now been simplified to remove all dependence on the line formation itself and instead is solely the temperature dependence of the ionic equilibrium (i.e., the curves in the top panel of Figure \ref{Fig2}). The contribution function can be simplified in this way since the ratio of the ionic density ratios have already accounted for the incident radiation and resonant excitation physics -- a process that cannot be applied with traditional DEMs given the dependence of collisional excitation on density. 

\par 

\subsection{R-DEM Inversion Procedure}
\label{procedure}

In this work, we will infer the R-DEM with the \ion[Fe x], \ion[Fe xi], and \ion[Fe xiv] line observations. We thus apply Equation \ref{eqnFinalRDEM} to both the \ion[Fe xiv]/\ion[Fe xi] and \ion[Fe x]/\ion[Fe xi] ionic density ratios and fit a density distribution of the $RDEM(T_e)$ that satisfies both ratios simultaneously (note that the elemental abundance terms disappear since we are only using Fe lines). That is, we use \ion[Fe xi] as the reference density, scale the other densities by the inferred ionic density ratio, then solve for what R-DEM would best fit the two ratios of the integrals (i.e., Equation \ref{eqnFinalRDEM}). Specifically, we use 201 bins of temperature ranging from 0.6 to 2.6 MK, with a spacing of 0.01 MK, {and use the standard Python scipy {\fontfamily{pcr}\selectfont optimize.minimize}\footnote{\url{https://docs.scipy.org/doc/scipy/reference/generated/scipy.optimize.minimize.html\#scipy.optimize.minimize}} function to fit the distribution of density coefficients. This optimization tool implements a Broyden–Fletcher–Goldfarb–Shanno algorithm, which is an iterative quasi-Newton method (default for {\fontfamily{pcr}\selectfont minimize}). The optimization procedure stops when it reaches the relative minimum of the difference between the inferred R-DEM and the observed ionic density ratios. 
\par
We initialized the temperature distribution as a uniform temperature distribution across the entire range from 0.6 to 2.6 MK. The function then varies the coefficients of each R-DEM bin to attempt to fit Equation \ref{eqnFinalRDEM} for both the \ion[Fe xiv]/\ion[Fe xi] and \ion[Fe x]/\ion[Fe xi] ionic density ratios simultaneously. Typically the procedure terminated after about 100 iterations, with a maximum number of iterations at 310 for the eclipse data and 445 for the MHD modeled lines (we set the maximum number of iterations to 2000 to prevent premature termination). The resulting distribution provides the R-DEM for each LOS. A simple fit is sufficient for this R-DEM inversion since the temperature response functionality of the bandpasses shown in the top panel of Figure \ref{Fig2} is simple. Consequently, the combination of bandpass integrals quickly converges on a solution -- even if only using these two ionic density ratios.
%eclipse 0 100.0 310
%psi  29 111.0 445
}

\par
As with all optically thin plasmas, this inversion {determines the integral} over all structures {along} the LOS which may well have different temperature distributions. The final inverted R-DEM should be thought of as the density-weighted temperature distribution along the entire LOS rather than for a well-defined volume of plasma.
\par
In this inversion, we are not taking into account any geometric LOS or limb-darkening effects on the radiative excitation of the lines. Along each LOS, the emission lines will experience similar dynamics from changes in the size and apparent horizon of the extended solar disk. Thus, most of these effects are removed by considering the ratio of the lines independently for each LOS, unless there is a dramatic asymmetry in the distribution of plasma temperatures along the LOS. There will be slight differences in the limb-darkening profile at different wavelengths, but these changes should only lead to small differences in the final line emission. As described in Papers I and II, we did account for limb-darkening when calibrating the absolute brightness of the lines, so we have already removed the limb-darkening effects somewhat. Finally, any effects of this kind will be most pronounced in the lowest part of the corona, which is also where we have to make corrections for collisional excitation using the MHD model. Clearly, the lower corona (below $\approx$ 1.3 \Rs) is not as robustly probed by this particular R-DEM inversion compared to the outer corona. Since EUV DEMs already probe the lower corona regularly, the primary focus of this R-DEM inversion method is to infer the temperature distribution farther out in the corona, which is otherwise exceptionally difficult to measure. 

\section{Results}

\subsection{Eclipse R-DEM}
\label{EclipseRDEM}

\begin{figure*}[t!]
\centering
\includegraphics[width =  6.5 in]{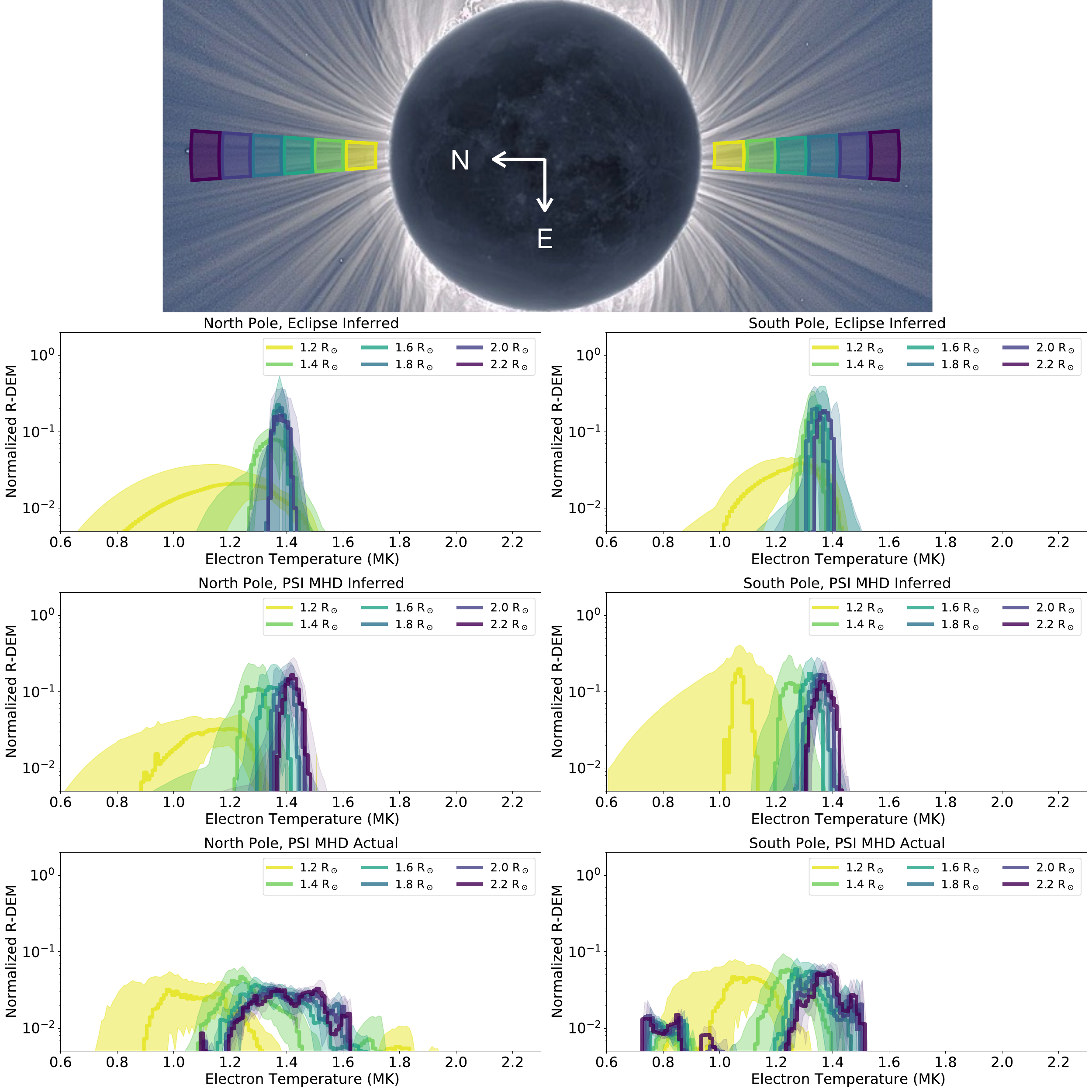}
\caption{Collection of R-DEMs for the coronal holes at the poles of the Sun. The top panel contains the white light image of the corona with a set of windows identified (8 degrees by 0.18 \Rs, steps of 0.2 \Rs). The bottom panels show the R-DEM results for the North Pole (left) and South Pole (right) where the top panels are from the eclipse data (see Section \ref{EclipseRDEM}), the middle panels are from the MHD forward modeled lines (see Section \ref{PSI}), while the bottom panels are the actual R-DEMs in the MHD model (see Section \ref{TestRDEM}). Every R-DEM inside each window is shown as a bold line, while the variance of the R-DEM distributions inside the window is shown as the shaded bands around each R-DEM curve. \vspace{2mm}} %Note, the white light image has been rotated 90 degrees counter-clockwise.
\label{Fig3}
\end{figure*}

\begin{figure*}[t]
\centering
\includegraphics[width = 6.5 in]{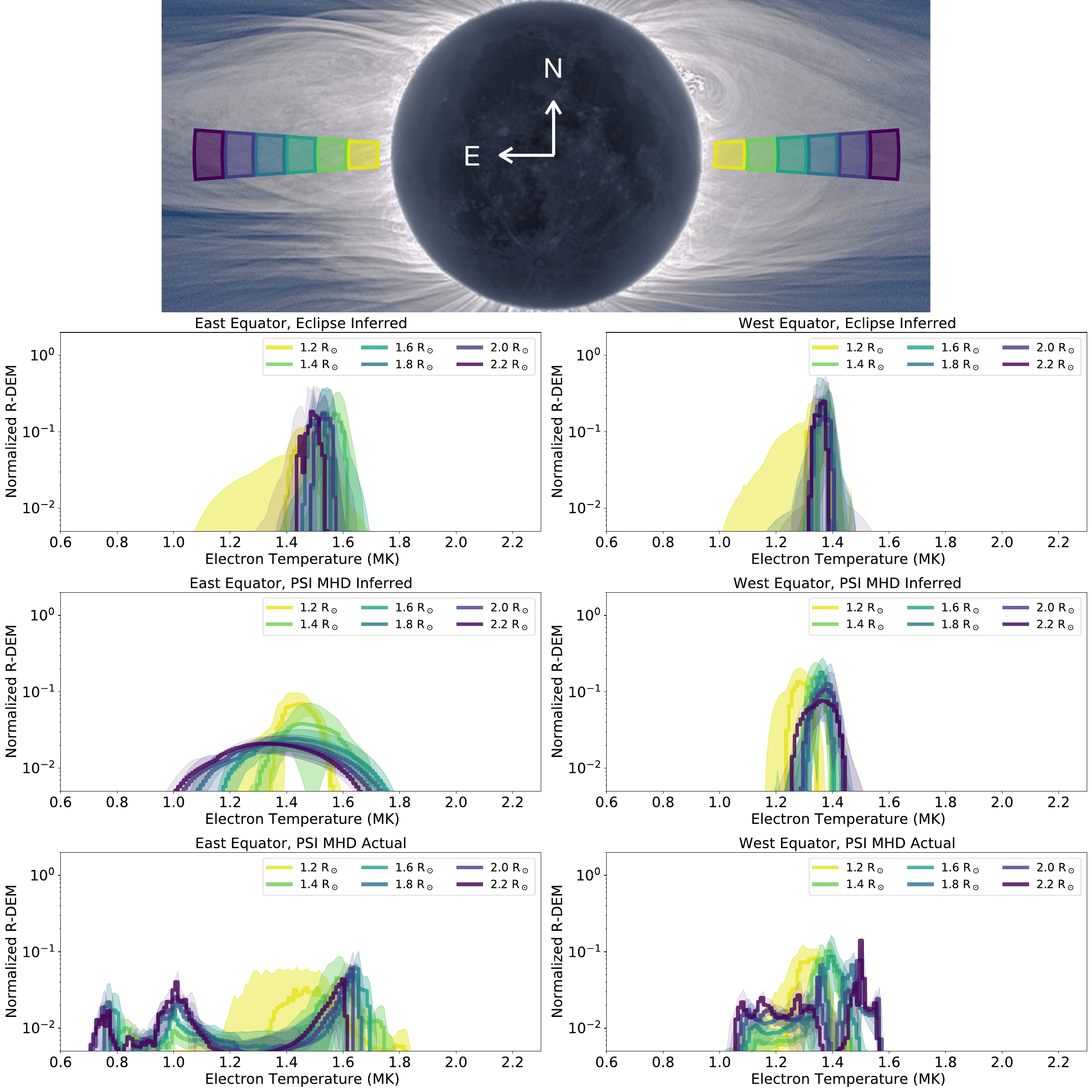}
\caption{Same as Figure \ref{Fig3}, except for windows centered on the streamers at the East (left) and West (right) equator. The white light image has the same orientation as Figure \ref{Fig1}. \vspace{2mm}} 
\label{Fig4}
\end{figure*}

A collection of $RDEM(T_e)$ inferences for the {observed emission} lines (see Section \ref{Eclipse}) is shown in the top panels of Figures \ref{Fig3} and \ref{Fig4} for samples from the poles and equator respectively. Each panel contains multiple sets of R-DEMs at different helioprojective distances ranging from 1.2 to 2.2 \Rs, in steps of 0.2 \Rs, inside of an 8-degree by 0.18 \Rs \ window centered on the various latitudinal regions. The R-DEM curves are colored according to the distance range they were taken at, as shown in the legend. Every individual R-DEM inversion result inside each distance range and latitudinal direction is shown as faint lines, while the median average of all R-DEMs in the window is shown as bold lines in the figure. 

%
%\begin{figure*}[t!]
%\centering
%\includegraphics[width =  6.5 in]{Fig5.pdf}
%\caption{Same as Figures \ref{Fig3} and \ref{Fig4}, except for windows centered roughly on the North (left) and South (right) edges of the Eastern streamer at position angles of 60 and 110 degrees respectively. Note, the white light image has been rotated 90 degrees counter-clockwise then flipped along the North-South axis to keep format consistent with the earlier figures.} %}
%\label{Fig5}
%\end{figure*}
%
%\begin{figure*}[t!]
%\centering
%\includegraphics[width = 6.5 in]{Fig6.pdf}
%\caption{Same as Figures \ref{Fig3} -- \ref{Fig5}, except for windows centered roughly on the North (left) and South (right) edges of the Western streamer at position angles of 300 and 250 degrees respectively.  Note, the white light image has been rotated 90 degrees counter-clockwise, the same orientation as Figure \ref{Fig3}.}
%\label{Fig6}
%\end{figure*}

\par
Each R-DEM is normalized by the integral of the R-DEM overall $T_e$ bins to show the relative R-DEM distribution along the LOS rather than the density of the corona. The coronal density drops rapidly with distance, so for visualization, it makes more sense to present the results in this manner. These R-DEMs could be scaled by an inferred electron density of each region given a coronal density inversion using the polarized or total brightness of the K corona (a K corona map is shown in Figure \ref{Fig1}). However, we are not concerned with the absolute density of the corona here, rather we are interested in the relative $T_e$ distribution throughout the corona.

\par
\begin{figure*}[t!]
\centering
\includegraphics[width = \textwidth]{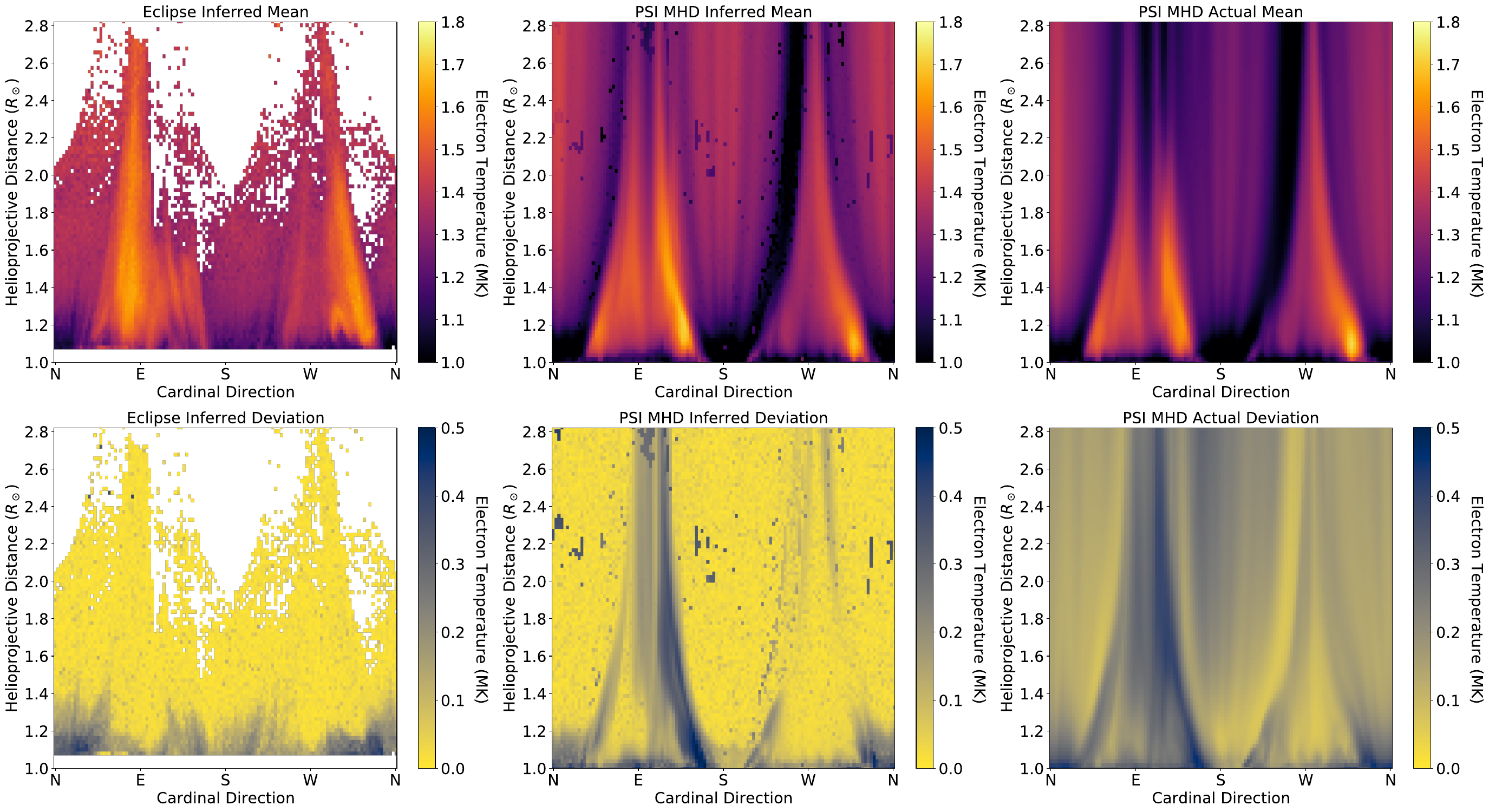}
\caption{Top left: Mean $T_e$ from the R-DEM inversion from the eclipse lines. Bottom left: Standard deviation of the R-DEM around the mean $T_e$. Middle: Same as left, for the MHD forward modeled R-DEM inversion. Right: Same as left, for the actual R-DEM distribution in the MHD model. \vspace{2mm}}
\label{Fig5}
\end{figure*}
\begin{figure*}[t!]
\centering
\includegraphics[width = \textwidth]{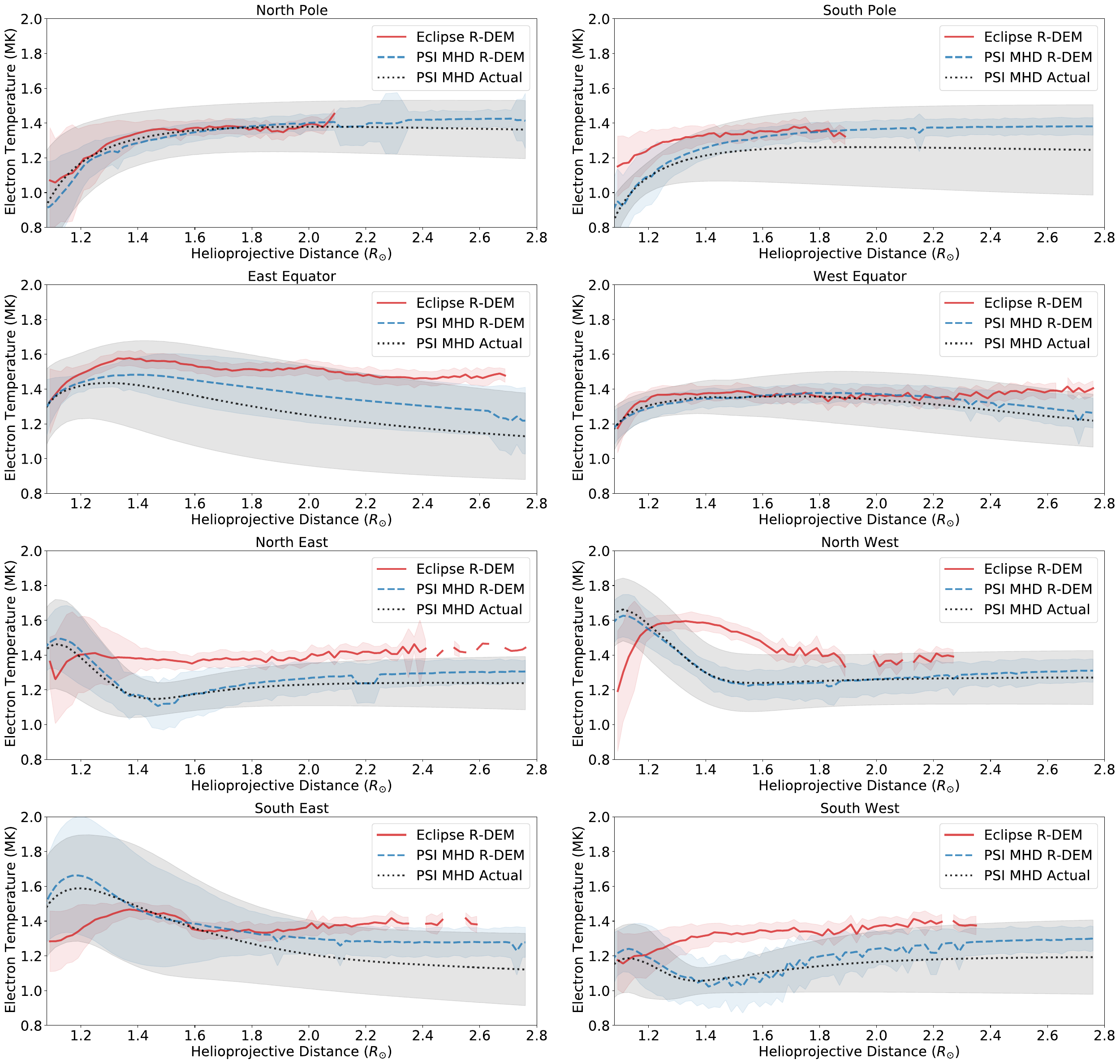}
\caption{Radial traces of the R-DEMs from the eclipse data (solid red line), inferred from the MHD model (dashed blue line), and the actual MHD model (dotted black line). The traces are taken from the median average inside a 15 degree wedge centered on the cardinal direction indicated in the title of each panel. The filled bands show the 1-$\sigma$ standard deviation of the R-DEM distributions within the wedge. \vspace{2mm}}
\label{Fig6}
\end{figure*}
\begin{figure*}[t!]
\centering
\includegraphics[width = 6.5 in]{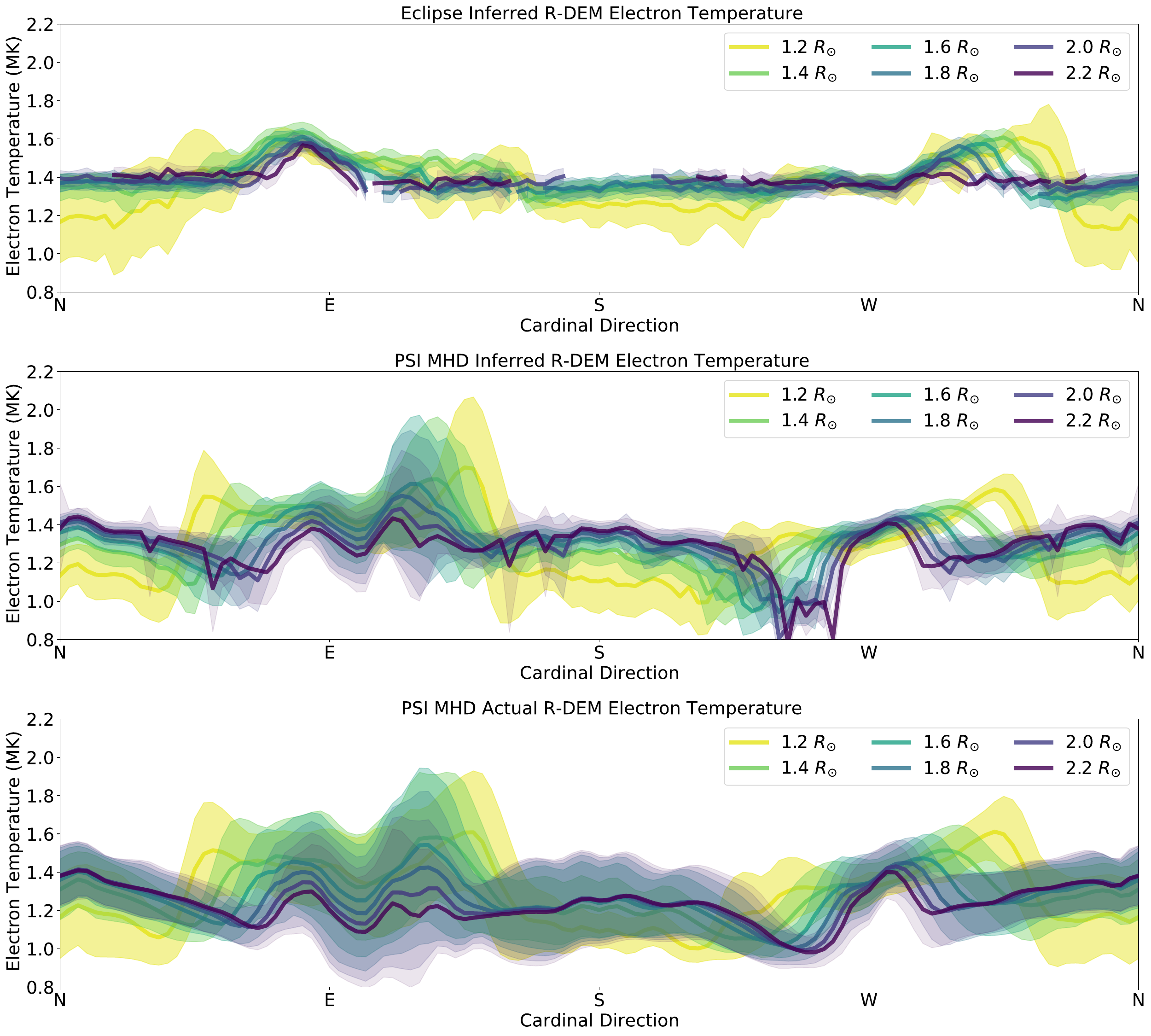}
\caption{Latitudinal traces of the R-DEM median (solid lines) and standard deviation (shaded bands) at a selection of distances from the eclipse data (top panel), inferred from the MHD model (middle panel), and the actual R-DEM in the MHD model (bottom panel). The traces are taken as the median average R-DEM at a set of fixed helioprojective distances within a band of 0.1 \Rs \ of each distance (i.e., $\pm 0.05$ \Rs). \vspace{2mm}}
\label{Fig7}
\end{figure*}

To create a representative map of the R-DEM distributions, we took the weighted mean of the R-DEM from each LOS to infer an average $T_e$, and the standard deviation (i.e., width) of the $RDEM(T_e)$ distribution around the mean. The resulting maps are shown in the left panels of Figure \ref{Fig5}. We then show a series of radial traces of the R-DEM at different latitudes in Figure \ref{Fig6}, and latitudinal traces at a series of radial distances in Figure \ref{Fig7}.

{
We find that the electron temperature generally varies from about 1.1 to 1.4 MK in the coronal holes, while the streamers are closer to 1.4 to 1.65 MK. The overall $T_e$ values in the coronal holes rise considerably from around 1.1 MK at 1.1 \Rs, up to almost 1.4 MK by about 1.4 \Rs. The streamers have a similar behavior, where they also rise in $T_e$ up to about 1.4 \Rs. On the POS, the streamers have considerable spatial variation, with multiple distinguishable stalks, whereas the coronal holes have much less spatial variation. Below 1.4 \Rs, there are a number of small cool plumes that extend into what appear to be closed field lines at the base of both the streamers and coronal holes. 
\par
The cores of the streamer have the highest $T_e$, especially in the eastern streamer, with values up to 1.65 MK. The western streamer is cooler in general, other than a closed loop and smaller stalk on the northwest side. The western streamer has a gap in the middle where the temperature is almost indistinguishable from the coronal holes, despite this region having higher electron densities (i.e., see the white light image and K corona map in Figure \ref{Fig1}). The width of the streamers also decreases considerably with height, occupying more than half of the corona at 1.2 \Rs, but becoming only a very small portion by 2.6 \Rs.
\par
The width of the R-DEMs along each LOS indicates a large spread in the $T_e$ throughout the corona (roughly $\pm$ 0.1 to 0.3 MK) below about 1.4 \Rs. However, beyond that distance, both the coronal holes and streamers appear almost entirely isothermal along the LOS with R-DEM widths less than $\pm$ 0.05 MK (see Section \ref{Isothermal}).
}

%\newpage
\subsection{PSI MHD R-DEM}
\label{PSI}
Next, we test the effectiveness of the R-DEM procedure with the state-of-the-art Magnetohydrodynamic (MHD) simulation of the PSI/MAS model prediction \citep{Mikic2018} of this eclipse. Papers I and II used this MHD simulation to forward model the K corona and line emission respectively, and in both cases, the model was found to rather accurately predict the eclipse observations. The forward-modeled line emission can now be used to test the validity of the R-DEM approach by comparing the R-DEM inferred from the forward-modeled lines to the true LOS $T_e$ distribution directly from the model. Using the same procedure described in Section \ref{formalism}, examples of the R-DEM inversion results are shown in Figures \ref{Fig3} -- \ref{Fig7}. In the same figures, we include the actual R-DEM from the MHD model in order to test the inversion process. {In Appendix \ref{Append}, we further discuss the details of the R-DEM distribution inside the PSI MHD model.
\par
The inferred $T_e$ distribution from the MHD model is very similar to the eclipse R-DEM in general, although the model inferred R-DEM has somewhat more variation in $T_e$ throughout the POS and along the LOS. The eastern streamer in particular has a large $T_e$ variance on the southern edge which persists beyond at least 2.8 \Rs.}

\subsubsection{Testing the R-DEM Inversion}
\label{TestRDEM}

\begin{figure*}[t!]
\centering
\includegraphics[width = \textwidth]{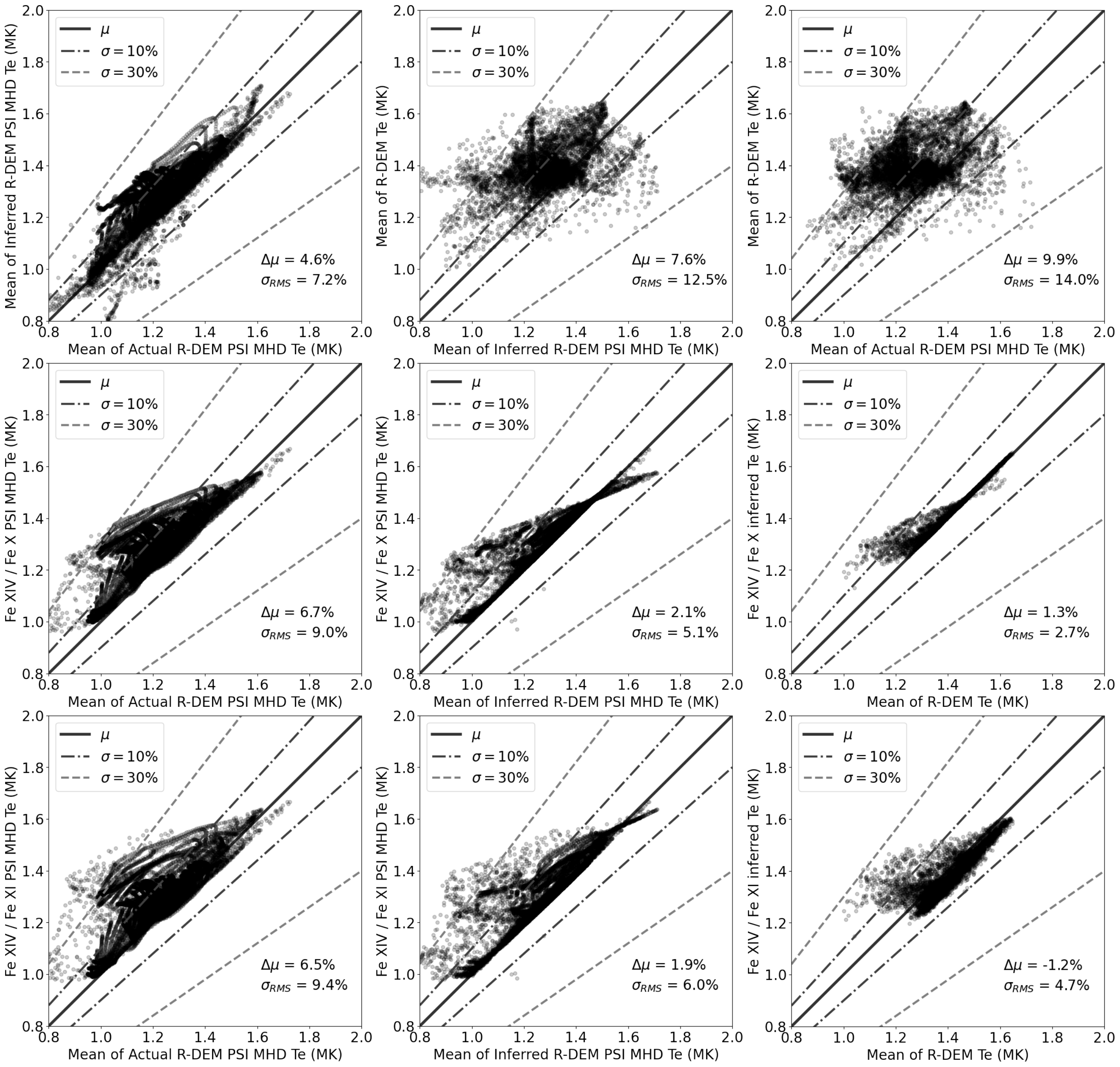}
\caption{Scatter plot comparisons between the various R-DEM and two line $T_e$ results from the eclipse, the PSI MHD forward-modeled lines and the actual LOS average of $T_e$ in the model. \vspace{2mm}}
\label{Fig8}
\end{figure*}

To test the new R-DEM methodology, we compare the results from the actual LOS $T_e$ distribution in the PSI MHD model with the inversion results from the forward-modeled lines. The actual R-DEMs, showcased in Figures \ref{Fig3} -- \ref{Fig7}, are quite similar to the inversion results, though the actual R-DEM shows substantially {wider $T_e$ distribution with more fine-scale structures {(see Appendix \ref{Append})}. Still, the average $T_e$ value is rather well recovered by the R-DEM inversion. A direct comparison between the average $T_e$ in the actual vs inferred model results is shown in the top left panel of Figure \ref{Fig8}, where the inversion result is $4.6 \%$ $\pm$ $7.2$ higher than the actual R-DEM distribution. Further, while the inversion tends to underestimate the width of the R-DEM distribution, it is able to recover realistic R-DEM widths when the actual R-DEM is wider than about $\pm 0.2$ MK. Along the southeast streamer edge, for example, the R-DEM inversion correctly finds widths between $\pm$ 0.3 and 0.4 MK. 
\par
It is not surprising that the inversion is unable to resolve the fine-scale $T_e$ structures and smaller R-DEM widths, since it only used three emission lines which have a somewhat broad temperature response compared to the shape of the actual R-DEM distribution, which reduces the precision that is possible to achieve. This effect is similar to the isothermal bias caused by two line-ratios \citep{Weber2005} and for some DEMs depending on the temperature response curves of the instrument \citep{Guennou2012a, Guennou2012b}. Future observations of additional lines such as \ion[Fe xiii] (1074.7 nm) would increase the temperature resolution of a similar R-DEM inversion. Thus, any inferred R-DEM widths below about 0.2 MK are not necessarily accurate in the inversions presented here. Instead, smaller widths should be considered as below that sensitivity limit.}
\par

\section{Discussion}
\label{disc}
In Section \ref{formalism}, we introduced the radiative analog of a DEM, called an R-DEM. We then applied this method in Section \ref{EclipseRDEM} to the \ion[Fe x], \ion[Fe xi], and \ion[Fe xiv] data from the 2019 TSE (see Section \ref{Eclipse}), as well as to the forward-modeled emission lines from the PSI MHD prediction for this eclipse in Section \ref{PSI}. {We compared the inferred model R-DEM distribution to the actual distribution of the MHD model in Section \ref{TestRDEM} to verify the accuracy of the inversion procedure.} Now, we will discuss the R-DEM inversion results from both the eclipse and model and compare them to the actual LOS $T_e$ distribution in the model.
\par

\newpage
{
\subsection{The $T_e$ of Coronal Holes and Streamers}
\label{temp}
%\subsubsection{Coronal Holes}
The novelty of this work is not the finding of any particularly unexpected $T_e$ values, but rather that we have inferred the $T_e$ values at a higher spatial resolution than has ever been achieved beyond the low corona (below 1.5 \Rs), and that we have inferred the LOS variance of $T_e$ (discussed in Section \ref{Isothermal}). EUV-based observations cannot constrain the electron temperature as far out as we have here, particularly in coronal holes where the signal is especially low. This high spatial resolution of the inference allows unique findings about both coronal holes and streamers. 

\par
We find the coronal holes to have electron temperatures of about 1.1 to 1.4 MK, where the values begin low, rise with helioprojective distance, then plateau at 1.4 MK by around 1.4 \Rs. These values are somewhat higher than the estimate of 1.0 to 1.3 MK (for the same helioprojective distances) by \cite{Habbal1993}, who reviewed a large number of earlier coronal hole $T_e$ inferences using X-ray, EUV, and white light data. These lower coronal hole $T_e$ values are also consistent with many years of in situ charge state data of coronal holes \citep{Smith2003, Habbal2010a}. One reason for this slight difference may not be due to actual differences in the electron temperature, but rather due to changes in the ionic equilibrium estimates used to infer $T_e$. For example, the Fe ions used in this work had their equilibrium curves shifted to a higher temperature by approximately 0.1 to 0.15 MK in the more recent estimates in CHIANTI compared to older, commonly used computations such as \cite{Arnaud1992}. These small changes in the ionic equilibrium estimates are sufficient to explain the small discrepancy between the coronal holes in this work compared to older results. In all cases, coronal holes are found to be rather isothermal (other than a slight increase in $T_e$ up to 1.4 \Rs) and consistent in the inferred $T_e$ values regardless of the inference method or period of the solar cycle, so it seems reasonable that improved ionic calculations are the only variable in this case.

\par
As for the streamers, we find considerable complexity in the structure of the $T_e$. Unlike the coronal holes, it is difficult to give a singular value of the average streamer temperature, as it varies significantly at high spatial resolution. Overall, the $T_e$ values in streamers range from about 1.4 to 1.65 MK, and we describe some of the fine-scale features in Section \ref{EclipseRDEM}. These $T_e$ values are relatively consistent with streamer observations from \cite{Raymond1997}, who used UVCS observations (mostly of collisional lines) to find a rough temperature of about 1.6 MK. However, they did not perform a full $T_e$ inversion. Indeed, \cite{Raymond1997} used lines from several different elements, so the observed brightness is convolved with the elemental abundance which will vary due to the FIP effect (see \citealt{Laming2015}). In this work, we used three lines of Fe which removes any FIP effect dependence on the $T_e$ inference. Similarly, \cite{Ko2002} found $T_e$ values near 1.5 MK in ``quiet-Sun" streamers and higher $T_e$ plasma above active regions (3 MK), but there were no active regions on the Sun during this particular eclipse. Additionally, \textit{Ulysses} ionic composition data yielded 1.3 to 1.8 MK near the solar equator at solar minimum \citep{Smith2003}, which can be assumed to have originated from coronal streamers similar to the ones seen in this eclipse. 

\par
The contrast between the east and west streamers is a great example of the importance of these observations, specifically in showcasing the diversity of streamer dynamics even near solar minimum when no active regions are present on the Sun. There are turbulent structures inside the eastern streamer visible in the white light image (top panel of Figure \ref{Fig1}, see \citealt{Habbal2021}) which correspond to the highest $T_e$ values anywhere in the corona during this eclipse. The western streamer, on the other hand, has a core with a much lower $T_e$ that is almost indistinguishable from the coronal holes despite having a significantly higher density. These very different streamer temperatures indicate that there are never truly ``quiet-Sun" streamers, but rather different levels of activity and dynamics that can create quite unique structures. \cite{Chitta2023} recently found similar fine-scale structures, complexity, and diversity of coronal streamers using white light and EUV observations along with MHD modeling -- albeit without any $T_e$ measurement.

\par

Both the coronal holes and streamers have a slight rise in $T_e$ from the base of the corona out to about 1.4 \Rs. The cause of this ubiquitous $T_e$ rise is not entirely clear, but there are two likely effects that could be contributing to the observed effect. One explanation might be a temperature-dependent scale height effect. \cite{Aschwanden2000} described how for an unresolved combination of hydrostatic loops, the higher temperature loops will have larger scale heights. Thus, higher temperature structures will have a more shallow density gradient and will increasingly dominate the emission at larger distances. The same principle should apply to hydrodynamic flows with different base temperatures as well. The other possibility is that the temperature rise may actually be a change in the $T_e$ of the plasma as it flows outward. Simulations of coronal heating often show a rise in temperature regardless of the exact heating mechanism used (e.g., \citealt{Habbal1995, Verdini2010, Matsumoto2014}). Additionally, as with all optically thin plasmas, there are LOS effects from overlapping structures which may change the inferred temperature depending on the exact structures present in the corona. Disentangling these effects is not a trivial endeavor and is beyond the scope of this work. Nevertheless, these observations provide constraints for benchmarking future coronal heating simulations.}

\subsection{On the Isothermality of the Middle Corona}
\label{Isothermal}
{
One notable attribute of the eclipse R-DEM inversion results is that there is a large temperature spread ($\pm$ 0.2 to 0.35 MK) in the lower corona (i.e. the yellow lines at 1.2 \Rs \ in Figures \ref{Fig3}, \ref{Fig4}, and \ref{Fig7}), with both much higher and lower $T_e$ values than found with the two line ratios (see Figure \ref{Fig1}). This spread then disappears at higher distances. In fact, all of the eclipse R-DEMs converge to an almost isothermal $T_e$ value (spread of $<0.05$ MK) by $\approx$ 1.4 \Rs, which is highlighted by the low $T_e$ standard deviation of the R-DEM (bottom left panel of Fig. \ref{Fig5}). 

\par
The R-DEM inversion from the model has a somewhat wider $T_e$ standard deviation throughout compared to the eclipse inversion, with the most dramatic spread occurring on the southern edge of the eastern streamer (see Figures \ref{Fig5} and \ref{Fig7}). As discussed in Section \ref{TestRDEM}, this three-line R-DEM inversion is not sensitive to distributions with widths less than about 0.2 MK but it is capable of detecting wider distributions. It is likely that if the actual distribution was significantly wider, then the R-DEM inversion should have detected it given its performance on the forward-modeled data. Our findings thus imply that in most of the corona, the emission weighted temperature is near isothermal (to at least $\pm 0.2$) beyond 1.4 \Rs, albeit with $T_e$ differences between coronal holes and streamers on the POS. However, it is possible that an unknown systematic error related to photometric noise or other optical issues may be suppressing the inferred R-DEM width, especially since we do not have a high $T_e$ precision with only three emission lines. Nevertheless, the isothermal nature of the corona appears at quite a low helioprojective distance where the line emission signals were strong and the MHD inversion results were able to recover wide R-DEMs in the eastern streamer out to 2.8 \Rs.

\par

The eclipse R-DEM inversion is indicating that (at the time of this eclipse) the corona below about 1.4 \Rs \ was a mixture of various structures along the LOS which have a wide range of temperatures (i.e., streamers and coronal hole plumes overlapping on the POS), which is expected near the Sun given the wide range of $T_e$ found in DEMs which have been achieved with EUV and X-ray observations (e.g. \citealt{Pottasch1964, Morgan2017}). There could also be scale height effects contributing to this inference, as discussed in Section \ref{temp}. 
\par
Beyond 1.4 \Rs, the outer corona becomes an almost isothermal plasma at the root of the solar wind. This finding supports the recent work of \cite{Habbal2010a, Habbal2021}, who found similar behavior in the corona and solar wind regardless of the phase of the solar cycle. We do find some small $T_e$ standard deviation between the coronal structures, but that spread is entirely on the plane-of-sky (POS) rather than along each LOS, and the majority of the corona becomes increasingly dominated by the highly isothermal coronal holes as the corona transitions into the solar wind. 
\par
Similar findings were reported by \cite{DelZanna2023}, who performed a collection of DEM and two-line ratio $T_e$ inferences from data taken during this same eclipse. In particular, they found a rather isothermal $T_e$ in the western streamer at 1.08 \Rs, and a wider DEM at the same distance in the Eastern streamer. The \cite{DelZanna2023} analysis incorporated infrared spectra from AIR-Spec \citep{Samra2022a, Samra2022b}, an airborne spectrometer flown during the eclipse, finding rather consistent two-line $T_e$ values for a number of infrared and EUV lines, supporting the concept that the corona was largely isothermal. However, that study was limited in its spatial coverage and elongation compared to the results presented here. Future work should incorporate additional lines as well as data from different phases of the solar cycle to better explore the behavior of the coronal R-DEM.
}

\par

\subsection{Testing the PSI MHD Simulation}
\label{TestPSI}
Since the mean of the R-DEM inversion with the forward-modeled MHD lines is a close fit to the mean of actual temperature distribution in the model (see Section \ref{TestRDEM}), we can reliably use the R-DEM results from the eclipse data to benchmark the PSI MHD simulation itself. The top middle and right panels of Figure \ref{Fig8} show direct comparisons between the eclipse inferred R-DEM average $T_e$ with the model R-DEM inversion (top middle) and the actual $T_e$ mean in the model (top right). Both comparisons are remarkably similar, with the eclipse inferred mean being somewhat higher than the model mean for the majority of LOSs. Specifically, the eclipse inversion is $7.6 \%$ $\pm$ $12.5$ higher than the model inversion, and $9.9 \%$ $\pm$ $14.0$ higher than the actual model mean $T_e$. These comparisons indicate that the model is very close to reproducing the correct temperature distribution, with perhaps a slight underestimation of $T_e$ on average. {The model inferred R-DEM also has slightly more temperature variability throughout the corona than the eclipse data, which is essentially isothermal at each distance (see Figure \ref{Fig7}). The variability is especially notable in the coronal holes, where the model has a significant peak at the north pole (about 1.4 MK) compared to the minima in the northeast (1.1 MK) and southwest (1 MK). These small differences between the model and eclipse data are} consistent with Paper II, which found that the \ion[Fe xi] and \ion[Fe xiv] lines were brighter in the coronal holes than the model predicted as well. Those differences were interpreted as an under-heating of the polar magnetic field lines in the Wave Turbulence Driven heating model, or possibly some other limitations in the MHD modeling approach.

\par
{
The streamer core temperatures in the model inversion are generally close to the values in the eclipse data, however, the width and location of the streamer stalks are somewhat different. The eastern streamer in particular has a much large latitudinal extent in the model and has two distinct peaks of $T_e$ whereas the eclipse data only has one prominent peak with a few smaller stalks on the southern side. The model inversion also predicts that the eastern streamer $T_e$ should decrease considerably with distance from the Sun, whereas the eclipse inversion maintains a relatively similar temperature out to at least 2.2 \Rs. Both the model and the data show the width of the eastern streamer decreasing considerably at larger distances. 
\par
The MHD model makes even better predictions about the behavior of the western streamer, both in terms of the overall $T_e$ values, and a drift of the center of the streamer stalk from the northwest towards the south from 1.2 out to 2.2 \Rs. Similar to the eastern streamer, the model does somewhat overestimate the width of the western streamer. Even though the trends in the western streamer are similar between the model and data, the model has the precise latitudinal location of the streamer somewhat further south than in the eclipse data. This small difference in latitude may be explained by slight inaccuracies in the magnetic boundary conditions of the model, which incorporate synoptic magnetic measurements from the sun-earth line and estimates of the polar magnetic flux, both of which are imperfect proxies for the full-sun-surface flux-distribution at a given instant of time (e.g., \citealt{Riley2019}).}

\subsection{Comparison to Two-Line Inferences}
\label{TwoLine}
The R-DEM mean $T_e$ maps are rather similar to both the \ion[Fe xiv]/\ion[Fe x] and \ion[Fe xiv]/\ion[Fe xi] two-line ratio inferences of the $T_e$ (bottom right panels of Figure \ref{Fig1}) for both the eclipse data and MHD model. A collection of direct comparisons between both the model and eclipse two-line ratio inferences to their respective R-DEM means are shown in Figure \ref{Fig8}. The two-line ratio inferences from the forward-modeled lines match the MHD average temperature quite well, with \ion[Fe xiv]/\ion[Fe x] (\ion[Fe xiv]/\ion[Fe xi]) returning a $T_e$ value that is $6.7 \%$ $\pm$ $9.0$ ($6.5 \%$ $\pm$ $9.4 $) higher than the actual LOS average (bottom left panels). Similarly, the \ion[Fe xiv]/\ion[Fe x] (\ion[Fe xiv]/\ion[Fe xi]) ratio inference finds an average value that is $2.1 \%$ $\pm$ $5.1$ ($1.9 \%$ $\pm$ $6.0 $) higher than the full R-DEM inversion of the MHD model (bottom middle panels). In all cases, the average values are within the scatter of the collection of all LOS in the dataset. As for the eclipse observations, the \ion[Fe xiv]/\ion[Fe x] $T_e$ average is $1.3 \%$ $\pm$ $2.7$ higher than the mean R-DEM inversion values (middle right panel), whereas the \ion[Fe xiv]/\ion[Fe xi] $T_e$ average is $1.2 \%$ $\pm$ $4.7$ lower than the R-DEM mean (bottom right panel).
\par
Based on these comparisons, it is clear that the R-DEM inversion is able to better reproduce the actual LOS average temperature than the two-line ratios, but the simple two-line ratios are accurate to better than $10 \%$. So older eclipse data with only two lines can still be reliably used to infer the average LOS $T_e$ (e.g., \citealt{Boe2020a}). Still, there are tails of points at lower $T_e$ in the two-line comparisons that diverge from the R-DEMs (i.e., the left side of the distributions in the middle and bottom panels of Figure \ref{Fig8}), which correspond to the low regions of the corona that had a high standard deviation of the R-DEM $T_e$ distribution. These low helioprojective regions are very likely a mix of colder open-field regions and hotter closed-field regions overlapping on the LOS -- as illustrated by the R-EM of the PSI MHD model (see Figure \ref{Fig9} in Appendix \ref{Append}). These more complicated regions are not well resolved by a simple two-line average. Fortunately, this effect is rather confined to the lower corona (below 1.3 \Rs). These lower regions of the corona are also where there is a considerable amount of collisional excitation in these lines. We isolated the radiative component of the observed brightness based on the collisional excitation fraction predicted by the MHD model (see Section \ref{EclipseRDEM}; Paper II), but there could be uncertainties or systematic errors induced by this procedure. Future work should include observations of collisionally excited lines, such as the \ion[Fe xiii] 1079.8 nm line (in addition to the radiative \ion[Fe xiii] 1074.7 nm line), which could be used to observationally isolate the collisional excitation component without the need for the MHD model to perform the R-DEM inversion.

%\newpage
\section{Conclusions}
\label{conc}
We have introduced the analog of a Differential Emission Measure for radiatively excited visible emission lines in the corona, which we call an R-DEM. The R-DEM methodology required a slight modification from traditional DEMs made with collisionally excited lines in the EUV and X-rays (see Section \ref{intro}), which we detailed in Section \ref{RDEM}.
\par
We then applied the new R-DEM method to data from the 2019 total solar eclipse observations of \ion[Fe x], \ion[Fe xi], and \ion[Fe xiv] (see Section \ref{Eclipse}) as well as to the same forward-modeled lines from the PSI MHD model for this eclipse (see Section \ref{PSI}). We then compared the actual LOS distribution of $T_e$ in the model to the inferred R-DEM, finding that the R-DEM method with only three emission lines was able to accurately reproduce the average $T_e$ values in the model (see Section \ref{TestRDEM}). Moreover, the R-DEM method was able to recover the width of the distribution in regions of the PSI MHD model that had a LOS $T_e$ standard deviation greater than about 0.2 MK but was unable to distinguish the $T_e$ distribution over finer scales (see Section \ref{Isothermal}). Likely, this effect is caused by the broad $T_e$ responses of these lines, and the limit of only three used in the inversion. Future studies should incorporate additional lines to improve the temperature resolution of the R-DEM.
\par
Given the demonstrated ability of this three-line R-DEM inversion to recover wide $T_e$ distributions, it is interesting that the eclipse R-DEM results indicate that most of the corona has a very small standard deviation of the LOS $T_e$. In the low corona, below about 1.3 \Rs, there are wide distributions in the $T_e$ values, but beyond that distance, the corona seems effectively isothermal. The coronal hole plasma is remarkably isothermal at about {1.1 to 1.4 MK, while the quiescent streamers at the equator are closer to 1.4 to 1.65 MK}. There is clearly variation in the temperature between the quiescent streamers and the coronal holes, but the LOS variation of each structure is negligible. 
\par
This finding of near-isothermality supports the work of \cite{Habbal2010a, Habbal2021}, who found that both the coronal and solar wind plasmas are strongly weighted to the charge state of \ion[Fe xi], which corresponds to about 1.3 MK. These results are also broadly consistent with \cite{DelZanna2023}, who performed DEM and two-line $T_e$ inferences during the same eclipse with EUV and airborne Infrared observations \citep{Samra2022a}. These consistent findings provide strong constraints for models of the solar wind. Specifically, that relatively isothermal temperatures can be used for solar wind model initialization outside of the lower corona {(above $\approx$ 1.4 \Rs)}. Combining solar wind models and in situ ionic composition data with this R-DEM inversion method, applied to observations of visible and infrared coronal emission lines, could have immense potential for advancing our understanding of coronal heating and the exact nature of the link between the corona and solar wind.

\par
The configuration of the corona during the 2019 eclipse, in particular, is an example of a near-perfect solar minimum corona. Future work could use this R-DEM method to test if there is a larger LOS $T_e$ standard deviation during periods closer to solar maximum, or if the corona remains essentially isothermal. R-DEM inversions would also be highly valuable when the corona has been disturbed by a coronal mass ejection. For example, \cite{Boe2020a} found changes in the LOS average $T_e$ due to a passing CME during the 2017 eclipse with only a 2-line $T_e$ inversion, but was unable to probe the LOS behavior with only two emission lines. Similar observations with three or more lines could be used to probe the complete temperature structure of an erupting CME in the middle corona (between 1.5 to 6 \Rs).

\par
The comparison between the R-DEM inversions of the eclipse and PSI MHD model also enabled a test of the model predictions. Similar tests had already been performed by \cite{Boe2021a} and \cite{Boe2022} for this specific simulation using the inferred K corona and two-line $T_e$ inferences respectively. In those papers as well as in this work (i.e., Section \ref{TestPSI}), the PSI MHD simulation is found to be an excellent match to the eclipse data. There is perhaps a very slight underestimation of the average $T_e$ values by the model, but the differences are typically less than $10 \%$. Still, these slight differences might point towards additional areas of improvement for coronal heating formulation in the MHD model. Further, the small discrepancies between the MHD model and eclipse data highlight the complexity of fine-scale structures in the corona that the model is not yet able to perfectly match. Regardless, the accuracy of the MHD simulation predictions is excellent.

\par
The two-line ratio $T_e$ inferences (from \citealt{Boe2022}) were then compared to the R-DEM inversions of both the eclipse and MHD model (see Section \ref{TwoLine}). In all cases, the inferred $T_e$ from both the \ion[Fe xiv]/\ion[Fe x] and \ion[Fe xiv]/\ion[Fe xi] ratios are able to rather accurately infer the correct average R-DEM $T_e$, except for lower regions of the corona (below 1.3 \Rs) where there is wider temperature distribution including a mixture of both colder ($<$1 MK) and hotter ($>$1.5 MK) plasmas along the LOS. Thus, observations with only \ion[Fe xiv] and either \ion[Fe x] or \ion[Fe xi] can still be reliably used to infer the average LOS temperature.

\par
As demonstrated in this work, visible and infrared emission lines are powerful tools for inferring the physical state of the corona out to at least 2.8 \Rs. In \cite{Boe2022}, the \ion[Fe xi] line was observed as far as 3.4 \Rs, a distance limited by the size of the detector rather than by a loss of signal. At future eclipses, instruments should be optimized to observe these emission lines to significantly larger helioprojective distances, which would be incredibly valuable for studying the formation and evolution of the solar wind in the corona, especially since these ions should freeze-in well below 5 \Rs \ \citep{Boe2018, Gilly2020}. It may be possible to measure visible line emission out to 10 \Rs \ during eclipses, which could be directly compared to spacecraft plasma measurements from future perihelia of the Parker Solar Probe \citep{Fox2016}. {Indeed, \cite{Antonucci2023} recently demonstrated with Metis on \textit{Solar Orbiter} that the UV \ion[H i] Ly $\alpha$ emission can be detected beyond 6 \Rs.} The eclipse-based observations thus continue to provide a key tool for studying the plasma properties of the corona and nascent solar wind.

\par
The R-DEM method introduced here could be applied beyond eclipse observations as well. The new data from the ground-based UCoMP coronagraph \citep{Tomczyk2021} in particular, would be well suited for an application of this R-DEM method. UCoMP can observe the same, and several additional, visible and near-infrared lines. With a wide selection of emission lines, the R-DEM method should be able to infer the $T_e$ distribution below $\approx 1.5$ \Rs. Additionally, the R-DEM method may be applied to the high spatial-resolution coronal observations by DKIST \citep{Rimmele2020}. Studies with DKIST and UCoMP in the lower corona could be used to explore the effect of collisional excitation on the R-DEM inversion method (see Section \ref{EclipseRDEM}), especially since both will observe the collisionally sensitive \ion[Fe xiii] doublet (see Section \ref{TwoLine}). While DKIST is not as well suited for studying the formation of the solar wind compared to the eclipse observations, it could infer the R-DEM at exceptionally high spatial resolution in prominence cavities, polar plumes, jets, CMEs, etc., which could be of high value to multiple areas of solar physics.

\par
Finally, recent demonstrations of the radiative self-excitation of the 17.1 nm and 19.3 nm EUV lines beyond 2 \Rs \ \citep{Seaton2021} suggest that future work could even apply this R-DEM technique to EUV lines farther out in the corona -- albeit after accounting for the complicated resonant scattering process driven by the collisional excitation of the same line lower down in the corona. Comparisons between similar inversions using lines all across the electromagnetic spectrum would be valuable for constraining the temperature distribution of coronal plasmas at a wide range of heliocentric distances and even could be applied to collisionless astrophysical plasmas generally.

%%%%%%%%%%%%%%%%%%%%%%%%%%%%%%%%%%%%%%%%%%%%%%%%%%%%%%%%%%%%%%%%

%  ACKNOWLEDGMENTS
\begin{acknowledgments}
%\subsection*{Acknowledgments}
Observables presented in this paper and other eclipse data from our group can be found at: \url{https://www.ifa.hawaii.edu/SolarEclipseData/}. The PSI MHD model eclipse predictions can be found here: \url{https://www.predsci.com/corona/} 
\par
Financial support was provided to B. B. by the National Science Foundation under Award No. 2028173. S. H. and the 2019 eclipse expedition were supported under NASA grant NNX17AH69G and NSF grant AST-1733542 to the Institute for Astronomy of the University of Hawaii. C. D. was supported by the NASA Heliophysics Supporting Research and Living With a Star programs (grants 80NSSC18K1129 and 80NSSC20K0192). 
\end{acknowledgments}
\newpage
%\vspace{8mm}
\appendix 
%\vspace{-5mm}
\section{Actual R-DEM of the PSI MHD Model}
\label{Append}{

The PSI/MAS MHD model is resolved at exceptionally high spatial resolution compared to previous global models (see \citealt{Mikic2018}), and generally at a higher spatial resolution than the eclipse data presented in this work. Specifically, the latitude and longitudinal mesh spacing is 0.009 radians ($\approx$ 0.5 degrees) except within 20 degrees of the poles, where the mesh spacing is 0.0105 radians ($\approx$ 0.6 degrees). The radial mesh spacing is highly variable depending on the distance from the Sun, where the mesh spacing is less than 0.01 \Rs \ below 1.6 \Rs, $\approx$ 0.025 \Rs \ at 2 \Rs, and $\approx$ 0.05 \Rs \ at 3 \Rs. Consequently, the spatial resolution of the model is considerably finer than the eclipse data in most of the corona, which has mesh points of 0.02 \Rs \ by 3 degrees (in position angle). Further, each LOS integration through the model combines a multitude of bins so even when the model voxels are larger than the data resolution, their intersection probes the corona at a finer resolution. 

\par
For the purposes of comparing with the R-DEM inversions in this work, we integrated the actual R-DEM in the model with the same polar coordinate binning on the POS projection from the Earth's perspective and the same $T_e$ bins used for the R-DEM inversion, which have a width of 0.01 MK (see Section \ref{procedure}). Since the actual R-DEM was integrated over these bins, the resulting integrals are Radiative Emission Measures (R-EM; i.e., Equation \ref{REM}). A collection of these R-EM maps are shown in Figure \ref{Fig9} for a selection of temperatures from 0.8 to 1.9 MK. An animation showing every R-EM from 0.6 to 2.6 MK is available in the online version of the article. These R-EM maps show the amount of plasma at a given $T_e$ along each LOS, in units of $cm^{-2}$ (see Equation \ref{REM}). In this representation, the densities are not normalized for each LOS, as they were in Figures \ref{Fig3} and \ref{Fig4}. In Section \ref{TestRDEM}), we concluded that even this binned $T_e$ resolution is much higher than is possible to resolve using the three-line R-DEM used in this work, still the R-EM maps are interesting for revealing the finer details in the MHD simulation. 

\begin{figure*}[t!]
\centering
\includegraphics[width = \textwidth]{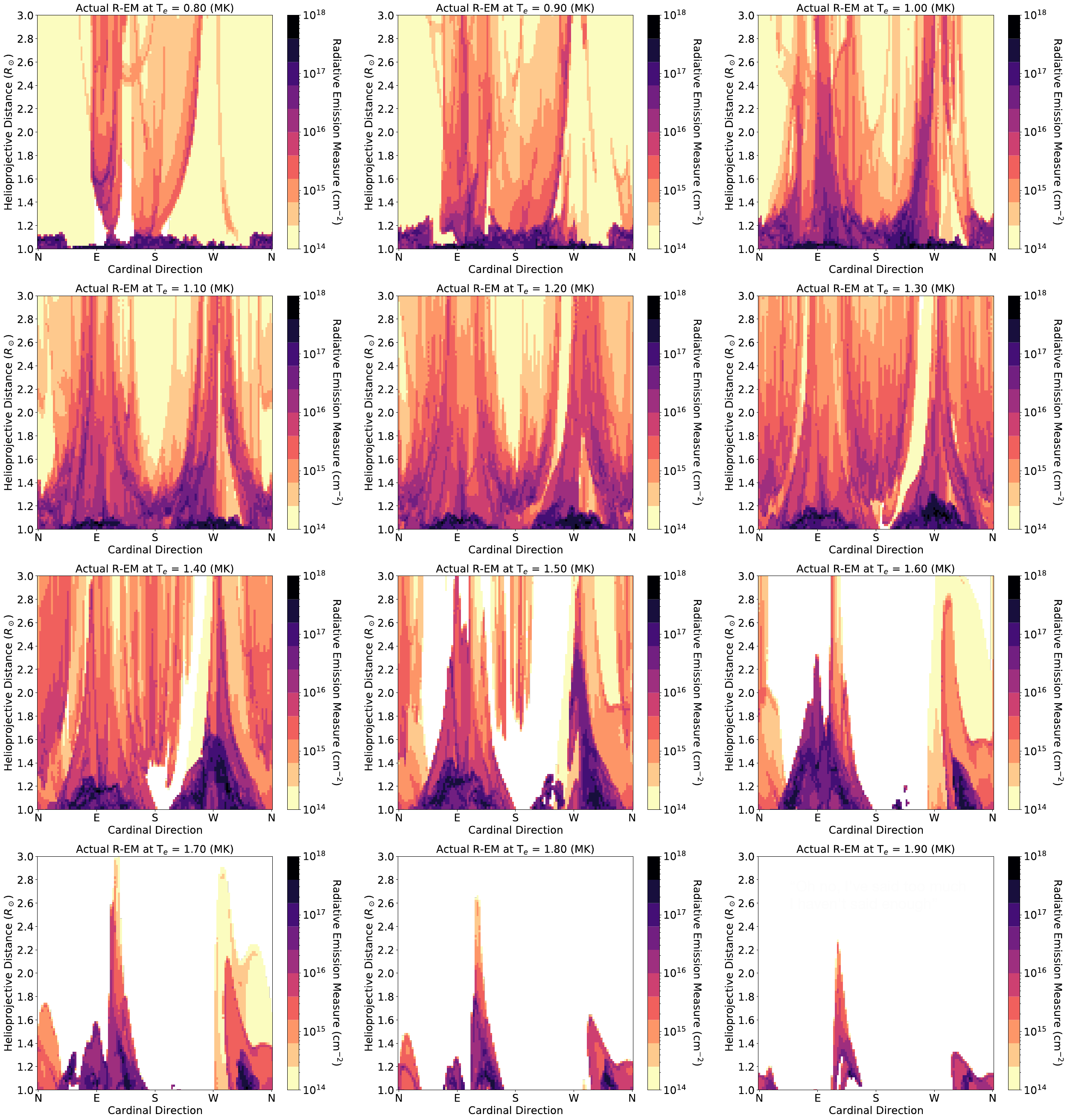}
\caption{Selection of radiative emission measures from the PSI MHD model ranging from 0.8 to 1.9 MK. Each panel shows the integrated R-EM over a range of 0.01 MK. An animation with every 0.01 MK slice of the R-EM from 0.6 to 2.6 MK is available in the online version of the article. \vspace{2mm}}
\label{Fig9}
\end{figure*}

\par

The model R-EMs show that there is a small density of plasma in the coronal holes at $T_e$ values below about 1 MK, especially the southern coronal hole, and at the base of the corona below 1.2 \Rs. Around 1 to 1.3 MK, the plasma density is somewhat uniformly distributed throughout the corona, roughly proportional to the bulk density. Above 1.5 MK, the plasma becomes increasingly confined to closed field lines in the streamers, and the extent of plasma at higher helioprojective distances decreases. The highest $T_e$ plasma (around 2 MK) in the south-eastern streamer overlaps with regions that had very low $T_e$ values as well. 
\par
In comparisons with the eclipse data, we found that the average $T_e$ in the model inversion is close to the eclipse data. However, the width of the model R-DEM inversion is considerably larger in some regions, particularly in the Eastern streamer. It is not clear why the model has a wider distribution of temperature in certain regions than the eclipse data indicates, but is likely related to the precise physical mechanism of the Wave-Turbulence-Driven (WTD) approach to heat the corona in the model \citep{Lionello2014, Downs2016, Mikic2018} and the inherent resolution of the global model grid, which naturally cannot resolve structures and contrasts below a certain size. To deduce the exact cause of these differences would require additional simulations where the precise assumptions and boundary conditions of the model are varied, which is beyond the scope of this work.

%Beyond these general trends in the R-EM distribution, there are an incredible number of fine-scale variations throughout the model R-EMs that showcase the complexity of the MHD modeled corona. Recent observational work has found similar complexity at fine-scale resolution throughout the corona with both white light eclipse imaging \citep{Boe2020b}, as well as deep exposures from space-based coronagraphs \citep{DeForest2018}, and even with EUV observations of Comet Lovejoy (C/2011 W3) as it passed through the corona \citep{Raymond2014}. This complexity in the corona also clearly persists in the solar wind, causing features such as switch-backs (e.g., \citealt{deWit2020}) and sharp Alfv\'enic spikes (e.g., \citealt{Horbury2020}). 
}

\bibliographystyle{apj}
\bibliography{2019RDEM}
\listofchanges
\end{document}